\documentstyle[twoside,11pt,amssymb,epsf]{article}

\catcode`\@=11
\long\def\@makefntext#1{
\protect\noindent \hbox to 3.2pt {\hskip-.9pt  
$^{{\tenrm\@thefnmark}}$\hfil}#1\hfill}			

\def\thefootnote{\fnsymbol{footnote}}
\def\@makefnmark{\hbox to 0pt{$^{\@thefnmark}$\hss}}	
	
\def\ps@myheadings{\let\@mkboth\@gobbletwo
\def\@oddhead{\hbox{}
\rightmark\hfil\tenrm\thepage}  
\def\@oddfoot{}\def\@evenhead{\tenrm\thepage\hfil
\leftmark\hbox{}}\def\@evenfoot{}
\def\sectionmark##1{}\def\subsectionmark##1{}}

\renewcommand{\thefootnote}{\fnsymbol{footnote}}

\newcounter{sectionc}\newcounter{subsectionc}\newcounter{subsubsectionc}
\renewcommand{\section}[1] {\vspace{25pt}\addtocounter{sectionc}{1} 
\setcounter{subsectionc}{0}\setcounter{subsubsectionc}{0}\noindent 
	{\twelvebf\thesectionc.\kern0.35cm #1}\par\vspace{8pt}}
\renewcommand{\subsection}[1] {\vspace{25pt}\addtocounter{subsectionc}{1} 
	\setcounter{subsubsectionc}{0}\noindent 
	{\twelvebf\thesectionc.\thesubsectionc\kern0.35cm #1}\par 
	\vspace{8pt}}
\renewcommand{\subsubsection}[1] {\vspace{25pt}\addtocounter{subsubsectionc}{1}
	\noindent
	{\twelverm\thesectionc.\thesubsectionc.\thesubsubsectionc\kern0.35cm 
	{\kern1pt\twelveit #1}}\par\vspace{8pt}}

\newcommand{\smalllineskip}{\baselineskip=11pt}

\def\ninecirc{
\begin{picture}(0,0)
\put(4.4,1.8){\circle{7.45}}
\end{picture}}
\def\ninecopyright{\ninecirc\kern2.75pt\hbox{\eightrm c}} 

\newcommand{\copyrightheading}[1]
	{\vspace*{-5cm}\baselineskip=11pt{\flushleft
	{\ninerm Fractals, #1}\\
	{\ninerm $\ninecopyright$\,\,\, World Scientific 
	 Publishing Company}\\
	 }}

\def\abstracts#1#2#3{{
	\centering{\begin{minipage}{5.0in}\tenrm\baselineskip=12pt
        \centerline{\twelvebf Abstract}
	\vspace{5pt}
	\parindent=0pc #1\par 
	\parindent=1pc #2\par
	\parindent=1pc #3
	\end{minipage}}\par}} 

\def\ARTICLES{\kern6.15cm\hbox{${\vcenter{\vbox{
	\hrule height 0.4pt width 4.562truein	    
	\hbox{\vrule width 0.4pt 		    
	height 0.6truein			    
	\raise0.565cm\hbox{\kern1pc
	\seventeenbf Articles}}
	\hrule height 0.4pt width 4.562truein}}}$}}  

\renewenvironment{thebibliography}[1]
	{\frenchspacing
	 \tenrm\baselineskip=12pt
	 \begin{list}{\arabic{enumi}.}
	{\usecounter{enumi}\setlength{\parsep}{0pt}
	 \setlength{\leftmargin 12.7pt}{\rightmargin 0pt} 
	 \setlength{\itemsep}{0pt} \settowidth
	{\labelwidth}{#1.}\sloppy}}{\end{list}}

\newcounter{itemlistc}
\newcounter{romanlistc}
\newcounter{alphlistc}
\newcounter{arabiclistc}

\newcommand{\fcaption}[1]{
        \refstepcounter{figure}
        \setbox\@tempboxa = \hbox{\footnotesize{\bf Fig.~\thefigure\phantom{00}}#1}
        \ifdim \wd\@tempboxa > 6in
           {\begin{center}
        \parbox{6in}{\footnotesize\smalllineskip{\bf Fig.~\thefigure\phantom{00}}#1}
            \end{center}}
        \else
             {\begin{center}
             {\footnotesize{\bf Fig.~\thefigure\phantom{00}}#1}
              \end{center}}
        \fi}

\newcommand{\tcaption}[1]{
        \refstepcounter{table}
        \setbox\@tempboxa = \hbox{\footnotesize\bf Table~\thetable\phantom{00}#1}
        \ifdim \wd\@tempboxa > 6in
           {\begin{center}
        \parbox{6in}{\footnotesize\smalllineskip\bf Table~\thetable\phantom{00}#1}
            \end{center}}
        \else
             {\begin{center}
             {\footnotesize\bf Table~\thetable\phantom{00}#1}
              \end{center}}
        \fi}

\def\@citex[#1]#2{\if@filesw\immediate\write\@auxout
	{\string\citation{#2}}\fi
\def\@citea{}\@cite{\@for\@citeb:=#2\do
	{\@citea\def\@citea{,}\@ifundefined
	{b@\@citeb}{{\bf ?}\@warning
	{Citation `\@citeb' on page \thepage \space undefined}}
	{\csname b@\@citeb\endcsname}}}{#1}}

\newif\if@cghi
\def\cite{\@cghitrue\@ifnextchar [{\@tempswatrue
	\@citex}{\@tempswafalse\@citex[]}}
\def\citelow{\@cghifalse\@ifnextchar [{\@tempswatrue
	\@citex}{\@tempswafalse\@citex[]}}
\def\@cite#1#2{{$\null^{#1}$\if@tempswa\typeout
	{IJCGA warning: optional citation argument 
	ignored: `#2'} \fi}}

\def\fnt#1#2{\footnotetext{\kern-.3em
	{$^{\mbox{\sevenrm #1}}$}{#2}}}

\def\runninghead#1#2{\protect\pagestyle{myheadings}
\markboth{\protect\nineit\,\,\,\,\,#1\hfill}
{\hfill\protect\nineit #2\,\,\,\,\,}}
\font\seventeenbf=cmbx10      scaled\magstep3

\font\twelverm=cmr10      scaled\magstep1
\font\twelveit=cmti10     scaled\magstep1
\font\twelvebf=cmbx10     scaled\magstep1

\font\tenrm=cmr10
\font\tenit=cmti10
\font\tenbf=cmbx10

\font\ninerm=cmr9
\font\nineit=cmti9

\font\eightrm=cmr8

\font\sevenrm=cmr7

\catcode`@=11					
\def\ps@plain{\let\@mkboth\@gobbletwo
     \def\@oddhead{}\def\@oddfoot{\ninerm\hfil\thepage
     \hfil}\def\@evenhead{}\let\@evenfoot\@oddfoot}

\def\ps@myheadings{\let\@mkboth\@gobbletwo	
\def\@oddhead{\hbox{}
\rightmark\hfil\ninerm\thepage}   
\def\@oddfoot{}\def\@evenhead{\ninerm\thepage\hfil
\leftmark\hbox{}}\def\@evenfoot{}
\def\sectionmark##1{}\def\subsectionmark##1{}}
\catcode`@=12

\begin{document}

\runninghead{Fast Relaxation Time In A Spin Model With Glassy Behavior}
{Fast Relaxation Time In A Spin Model With Glassy Behavior}

\renewcommand{\thefootnote}{\fnsymbol{footnote}}      

\thispagestyle{plain}
\setcounter{page}{1}

\copyrightheading{Vol.~0, No.~0 (0000)}

\vspace{6pc}
\leftline{\phantom{\ARTICLES}\hfill}

\vspace{3pc}
\leftline{\hskip-0.1cm\vbox{\hrule width6.99truein height0.15cm}\hfill}

\vspace{2pc}
\centerline{\seventeenbf FAST RELAXATION TIME IN A SPIN MODEL}
\baselineskip=20pt
\centerline{\seventeenbf WITH GLASSY BEHAVIOR}
\vspace{0.27truein}
\centerline{GIANCARLO FRANZESE}
\baselineskip=12.5pt
\centerline{\it Center for Polymer Studies and Department of Physics, Boston University}
\centerline{\it 590 Commonwealth Avenue, Boston, MA 02215, USA}
\vspace{0.36truein}
\abstracts{ 
We consider a frustrated spin model with a glassy dynamics characterized
by a slow component and a fast component in the relaxation process.
The slow process involves variables  
with critical behavior at finite temperature $T_p$ and has a
global character like the (structural)
$\alpha$-relaxation of glasses. 
The fast process has a more local character and can be associated to the 
$\beta$-relaxation of glasses.
At temperature $T> T_p$ the fast 
relaxation follows the non-Arrhenius behavior of the slow variables.
At $T\lesssim T_p$ the fast variables have an Arrhenius behavior,
resembling the $\alpha-\beta$ bifurcation of fragile glasses.
The model allows us to analyze the relation  between the dynamics
and the  thermodynamics.
}{}{}

\vspace{0.78truein}
\baselineskip=14.5pt
\section{INTRODUCTION}
\noindent
A recent review\cite{10q} has pointed out some of the 
leading questions in the theoretical study of glassforming liquids.
Following the fruitful path of theoretical studies on spin models with 
 glassy behavior\cite{Young}, here
we present results on a frustrated spin model that 
give an insight on some of these questions, showing relations between
the dynamics and the statics of the model.

We consider a model, introduced by
Coniglio and coworkers\cite{CdLMP,PSG,franzese0_FFdCC,breve,phil,PFF}, with two kinds of
variables, coupled to each other. One of them has a finite temperature
transition and a slow relaxation, while the other has a transition only
at zero temperature in two dimensions (2D) and has a fast relaxation at
finite temperature. The model has been show\cite{breve,PFF} to have
glassy behavior and the slow and fast components of the relaxation can be related to
the $\alpha$ and $\beta$ processes, respectively, of glasses. Here we show
numerical results resembling
the $\alpha$-$\beta$ {\it bifurcation process}\cite{10q}
observed in experiments on fragile liquids\cite{johari,stickel_wagner}.

In the following  we give an introduction to the problem.
In Sec. 2 we introduce the spin model, reviewing some results on its 
phase diagram and on its glassy behavior.
In Sec. 3 we show some numerical results about the
bifurcation in this model and the connection with the
thermodynamics.
In Sec. 4 we give the conclusions and the perspectives
for future work.

\subsection{Glasses And The $\alpha$-$\beta$ Bifurcation}
\noindent

The majority of liquids can form a glass if cooled at high enough rate.
The glass state is experimentally defined by measuring 
macroscopic quantities, like the diffusion coefficient or the
viscosity. The calorimetric glass transition temperature $T_g$ is
defined as the temperature at which the viscosity reaches a value of
$10^{12}$sPa and depends on the cooling rate. 

The glass
transition is considered as a dynamic  off-equilibrium phenomenon,
but a large amount of work has
recently shown that the dynamics of glassy systems is strictly related
to its static properties, i.e. to its thermodynamics\cite{pes}.
The main focus of these studies is the slow component of the dynamics.  
One of the aim of this paper is to show that also the fast processes can
be related to the thermodynamics of the system.

Following the Angell's classification\cite{Angell}, glassforming
liquids are divided in `strong' and `fragile'. 
Strong liquids are characterized by a relaxation time $\tau(T)$ that
increases with decreasing temperature $T$ in a way well described by an Arrhenius
law
\begin{equation}
\tau(T)=\tau_\infty \exp[A/(k_B T)]
\end{equation}
where $\tau_\infty$ is interpreted as a characteristic microscopic
relaxation time at infinite $T$
and  $A$ is
an activation energy (energy barrier) 
for global rearrangements ($k_B$ is the Boltzmann constant). 
A strong glassformer is, for example, SiO$_2$.

Fragile liquids are defined as those whose $\tau(T)$ shows a large
departure from the Arrhenius law (non-Arrhenius behavior) for $T>T_g$.
For $T<T_g$ the relaxation time is well described by an Arrhenius law.
Example of fragile glassformers are glycerol and orto-therphenyl.
A function widely used to fit data for fragile liquids is the
Vogel-Tamman-Fulcher (VTF) law 
\begin{equation}
\tau(T)=\tau_\infty \exp[B/(T-T_0)]
\label{VTF}
\end{equation}
where $\tau_\infty$ is a hight-$T$ characteristic time, $B$ and $T_0$
are fitting constants. Since $T_0<T_g$ the diverging behavior is never
reached. 

An alternative non-Arrhenius behavior is derived in the Mode Coupling
Theory (MCT)\cite{Gotze} for $T>T_g$, in which the relaxation time increases as a
power law of the temperature,
\begin{equation}
\tau(T)\simeq (T-T_{MCT})^{-\gamma} ~,
\label{MCT}
\end{equation}
where $T_{MCT}$ is the MCT transition temperature (with $T_{MCT}<T_g$) and
$\gamma$ is a parameter. 

The non-Arrhenius behavior can be explained, in a consistent way with
the MCT\cite{pes},  also by the Adam-Gibbs theory
of the excluded volume, in which the relaxation time is
\begin{equation}
\tau=\tau_\infty \exp(C/TS_c)
\label{AG}
\end{equation}
where $\tau_\infty\simeq 10^{-14}$s, $C$ is a constant and $S_c$ is the
entropy difference between liquid and crystal -- `excess' entropy.
Since $S_c$ is, usually, constant below $T_g$, the low-$T$ Arrhenius
behavior is recovered.

This phenomenology  concerns the structural $\alpha$-relaxation,
i.e. the process described by the long-time part of the relaxation
functions of the system.
For high $T$,
the relaxation functions, like the density-density correlation function,
have a simple exponential behavior (`simple liquid' regime) 
\begin{equation}
f(t)=f_0\exp(t/\tau) ~,
\label{exp}
\end{equation} 
with $f_0$ and $\tau$ depending on $T$ and where $t$ is the time.

Decreasing $T$ the relaxation functions usually show a `two-steps' behavior where 
a fast relaxation process, associated to the first step, is followed by
a slow relaxation process, the second step.
The second step is the $\alpha$-relaxation and is  associated to a macroscopic process
involving non-local 
degrees of freedom (as the global reorganization of the system)\cite{10q}. 
Below a characteristic temperature $T^*>T_g$, 
the second step is well approximated in most cases by the 
Kohlrausch-Williams-Watts (KWW) stretched exponential function 
\begin{equation}
f_{KWW}(t)=f_0\exp[(t/\tau)^\beta] ~,
\label{KWW_stret}
\end{equation} 
where $f_0$, $\tau$ and $\beta$ depend on $T$.
The KWW function  recovers the simple exponential for $\beta=1$,
with the parameter $\beta$ describing the departure from exponentiality.
The fit of glassy relaxation functions, usually, shows $\beta$
decreasing with the temperature, with $\beta<1$ for $T<T^*$.
In many cases, a better approximation of the second step
of the relaxation function is given by using 
the Ogielski\cite{Ogielski} stretched exponential function 
\begin{equation}
f_O(t)=t^{-x}f_{KWW}(t) ~,
\label{Ogielski_stret}
\end{equation} 
with $x$ depending on $T$.

The first step (or $\beta$-relaxation) is usually associated to any
microscopic process\cite{10q} occurring at very short time
scale. In the MCT for atomic liquids it is
associated to the fast diffusion of mobile particles inside `cages' of
immobile particles\cite{Gotze}. In molecular liquids it can be
associated also to the relaxation of internal degrees of freedom\cite{mossa}.

The $\beta$-relaxation presents some non-universal features in
fragile liquids\cite{johari}, and sometimes in intermediate 
liquids\cite{stickel_wagner}, that are not well understood. 
The $\beta$-relaxation time follows the non-Arrhenius
$\alpha$-relaxation behavior at high $T$ and shows a crossover 
(the $\alpha-\beta$ bifurcation) to an Arrhenius behavior at a
temperature $T_{\alpha-\beta}\simeq T_{MCT}$\cite{10q}.
To have an insight on this phenomenon and its connection with other
static and dynamic transitions, we consider the spin model presented in
the next section.

\begin{figure}
a) \mbox{ \epsfxsize=6cm \epsffile{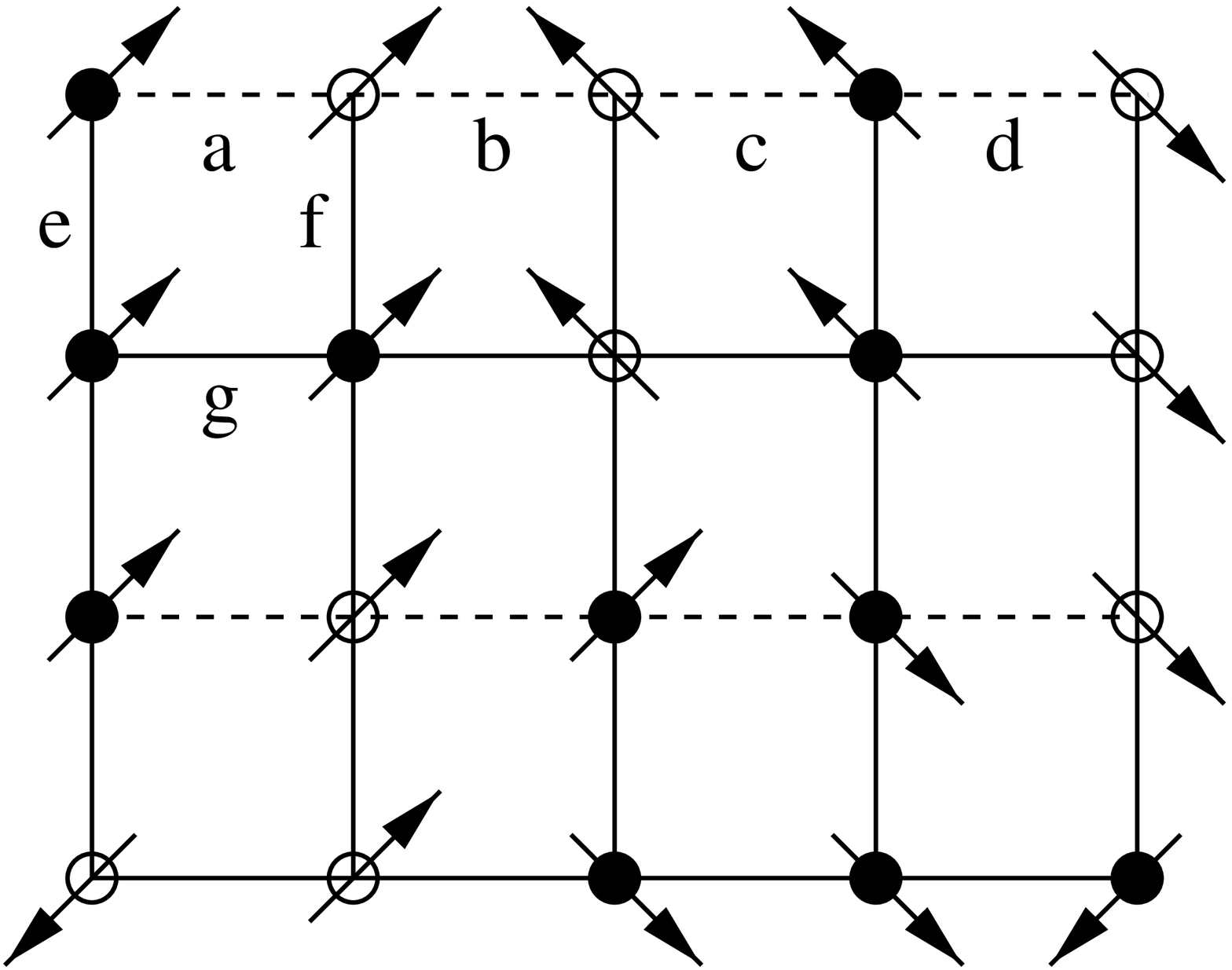} }
b) \mbox{ \epsfxsize=7cm \epsffile{ 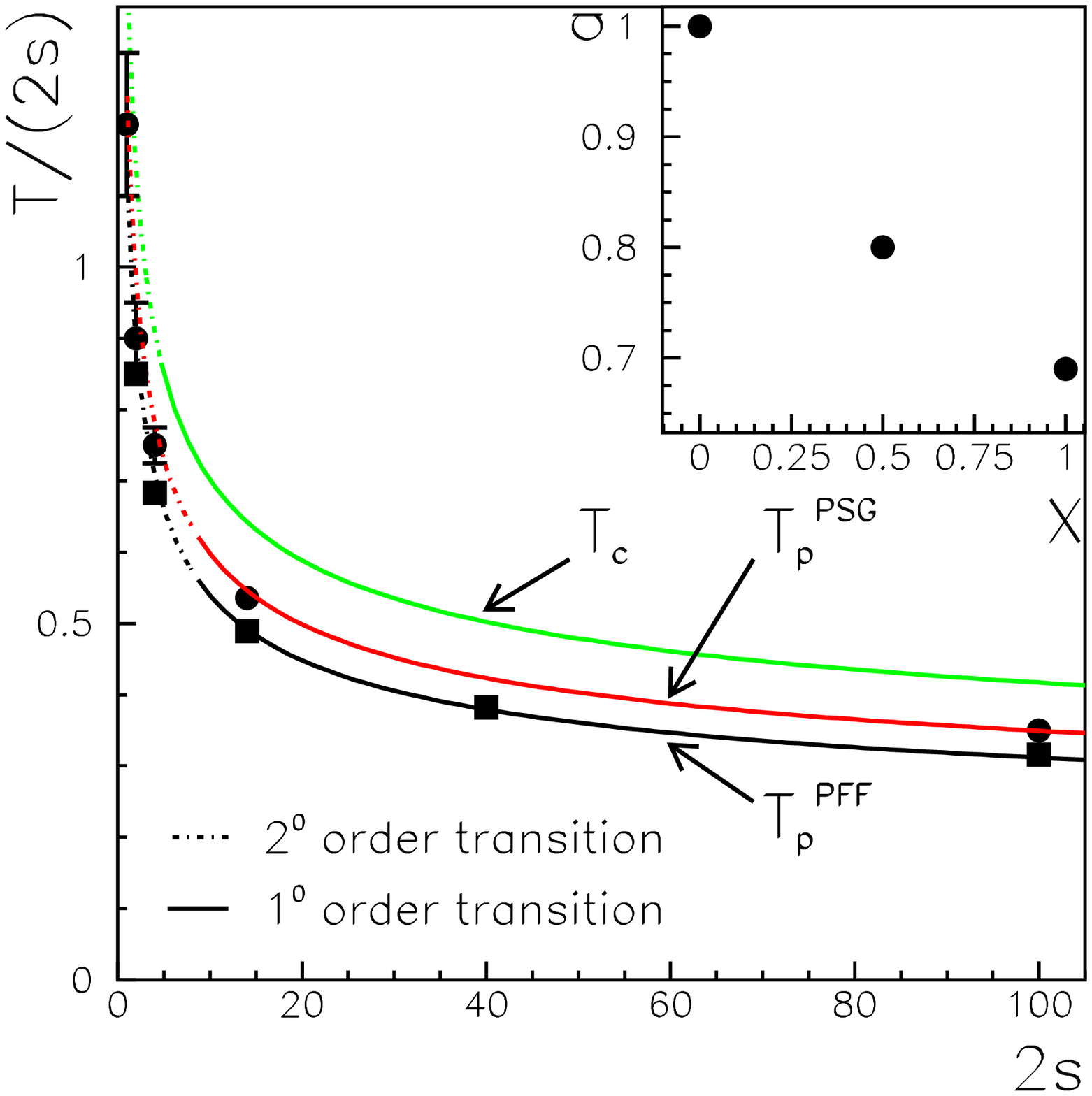 } }
\vspace{0.5cm}
\fcaption{a) A Potts fully frustrated (PFF) model on a square lattice: 
on each vertex there is variable $\tau_i\equiv S_i\sigma_i=\pm 1, \pm 2 \dots \pm s$ with
$s=4$. Here we represent the spin $S_i$ by an
open or a full dot (respectively positive and negative, for example)  
and the orientation $\sigma_i$ by an arrow pointing in 
4 different directions. 
Ferromagnetic (antiferromagnetic) interactions are represented by 
full (dotted) lines.
b) The analytic value of the transition temperature (in 2D)
$T_c \equiv T_p(q,0)$, with $q=2s$, 
of the ferromagnetic Potts model (upper line), corresponding to $X=0$,
is given together with the 2D Monte Carlo data for the
PSG model (circles), corresponding to $X=0.5$, 
with transition at  $T_p^{PSG}$ and for the 
PFF model (squares), corresponding to $X=1$, 
with transition at $T_p^{PFF}$.
The data are fitted with
Eq.(\protect\ref{TpX=0.5}) with
$T_p^{PSG} \equiv T_p(q,0.5)$ and
$T_p^{PFF} \equiv T_p(q,1)$.
The parameters $a(X)$ are shown in the inset. 
Errors are smaller than symbols size.
}
\label{lattice}
\end{figure}

\section{THE SPIN MODEL}
\noindent

We
consider a lattice model\cite{CdLMP}
with, on each lattice site, a 
Potts variable\cite{Wu} $\sigma_i$, with and integer number $s$ of states ($\sigma_i=1,\dots,s$), 
coupled to an Ising spin $S_i$ with two states ($S_i=\pm1$). 
The model can be considered as a schematic representation of 
structural glasses, such as  dense molecular glasses, plastic
crystal, or orto-therphenyl at low
temperature, 
with orientational degrees of freedom  frustrated
by geometrical 
hindrance between non-spherical molecules.
The orientational degrees of freedom are represented by the Potts
variables 
and the frustration is modeled by means of 
ferro/antiferromagnetic interactions for the Ising spins.
The model is defined by the Hamiltonian 
\begin{equation}
H_s\{\tau_i,\epsilon_{i,j}\}=
-2sJ\sum_{\langle i,j \rangle} \delta_{\epsilon_{i,j}\tau_i, \tau_j}
\label{hamiltonian2}
\end{equation}
where $\tau_i\equiv S_i\sigma_i=\pm 1, \pm 2, \dots , \pm s$ has 
$q\equiv 2s$ states,
the sum is extended over all the nearest neighbor  (n.n.) sites,
$J$ is the strength of interaction,
$\epsilon_{i,j}=\pm 1$ is a quenched variable that represents the sign
of the ferro/antiferromagnetic interaction, 
$\delta_{n,m}=0,1$ is a Kronecker delta.

In the original formulation\cite{CdLMP}, the Hamiltonian shows a clearer
separations between the two ($\sigma_i$ and $S_i$) coupled variables:
\begin{equation}
H_s\{S_i,\sigma_i,\epsilon_{i,j}\}
=-sJ\sum_{\langle i,j \rangle} \delta_{\sigma_i, \sigma_j}
(\epsilon_{i,j}S_i S_j+1) ~.
\label{hamiltonian}
\end{equation}
The Eq.(\ref{hamiltonian}) shows that the Ising spins interactions are 
diluted by a ferromagnetic Potts model, i.e. two n.n. Ising spins can
interact only when the corresponding n.n. Potts variables are in the
same state (or orientation). 

The model has been proposed in two versions. 
In the first  version, the interaction signs $\epsilon_{i,j}$ are randomly
assigned, giving rise to frustration
and disorder.  This model can be considered the generalization of an Ising
spin glass (SG) -- with two
states -- to a model with $2s$ states -- the Potts SG
(PSG)\cite{PSG}.
 
In the other version, the interactions signs are assigned in a
deterministic way and there is an odd number of $\epsilon_{i,j}=-1$
(antiferromagnetic interaction) on each lattice cell.
In this way the model has frustration and {\it no}
disorder and is the generalization of the Ising fully frustrated (IFF) 
model\cite{franzese0_FFdCC} to a $2s$-states Potts fully frustrated
(PFF) model\cite{PFF} (Fig.\ref{lattice}.a).
In the following we will review some results on the PSG and the PFF
model. 

\begin{figure}
a) \mbox{ \epsfxsize=6cm \epsffile{ 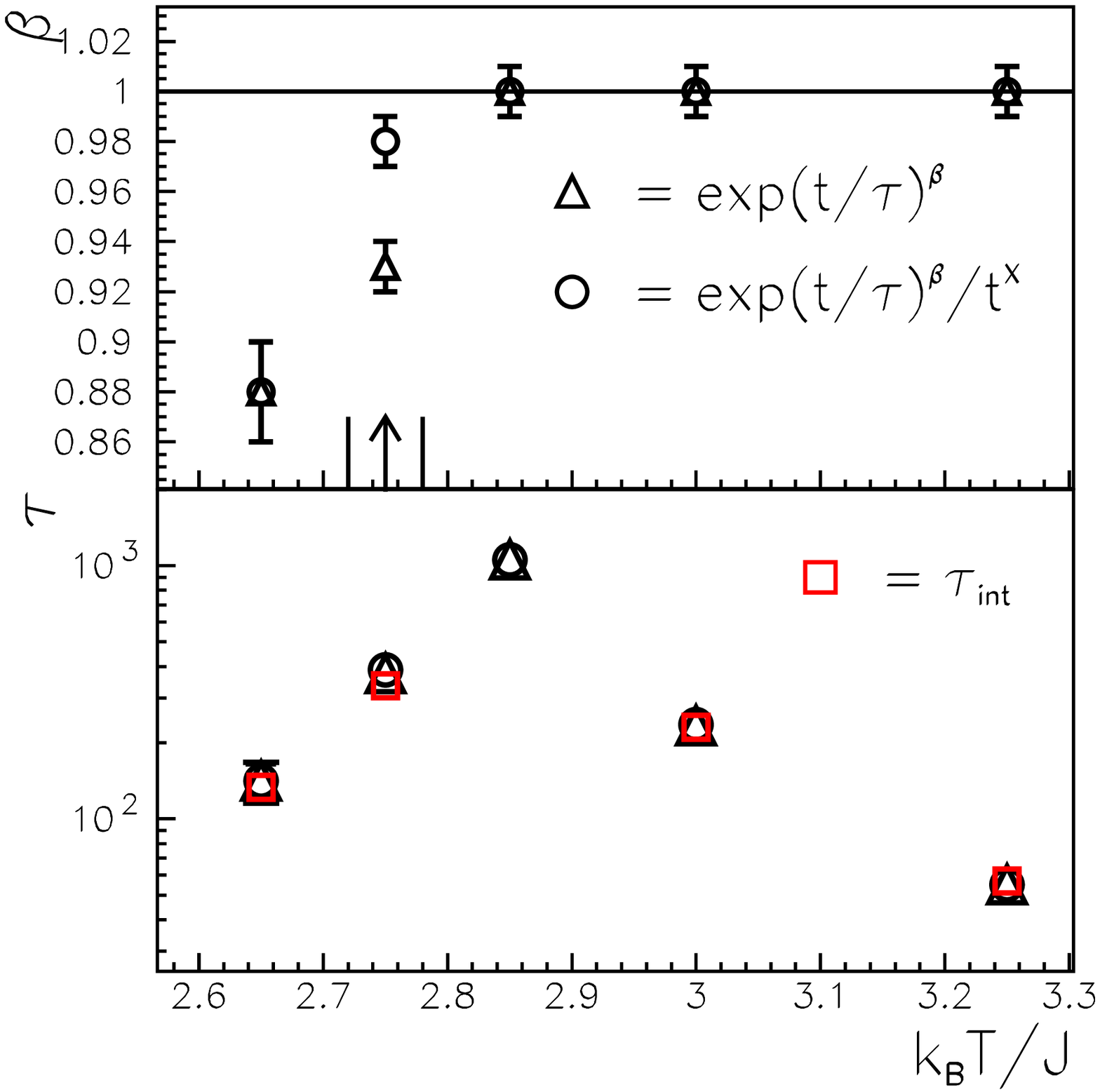 } } 
b) \mbox{ \epsfxsize=6cm \epsffile{ 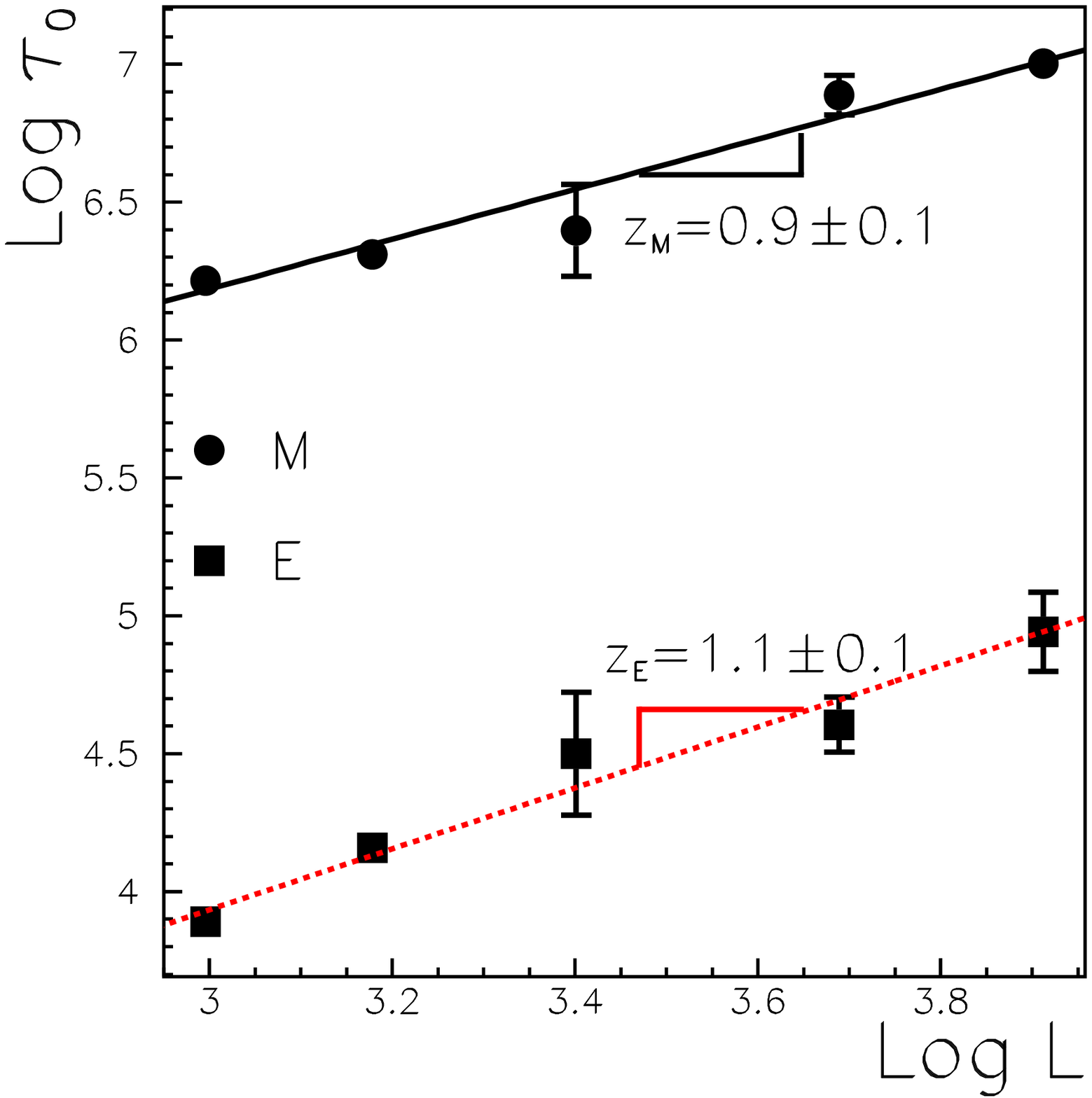 } }
\vspace{0.5cm}
\fcaption{The slow relaxation in the PFF model with $q\equiv 2s=4$ in 2D.
a) The fitting parameter $\beta$ (upper panel) and $\tau$ (lower panel)
for the second step of the correlation function of the Potts order parameter
$M$ (extrapolated to the thermodynamic limit). 
The circles are the parameters for 
the Ogielski stretched exponential form in
Eq.(\protect\ref{Ogielski_stret}), the triangles for 
KWW stretched exponential form in Eq.(\protect\ref{KWW_stret}).
The arrow shows the estimate of $T_p(4,1)$ in Eq.(\protect\ref{TpX=0.5}).
The lower panel includes also data for the integral
correlation time $\tau_{int}=\lim_{t_{max}\rightarrow
\infty}[\frac{1}{2}+\sum_{t=0}^{t_{max}}f_M(t)]$.
Where not shown, the errors are smaller that the
symbols size. 
b) 
The log-log plot of the autocorrelation time $\tau_0$, for the energy
density $E$ (squares) and the Potts order parameter $M$ (circles)
at the finite-size
transition temperature $T_p(L)$  as function of
the  size $L=20$, 24, 30, 40, 50.
The autocorrelation time $\tau_0$ is defined as the time (in unit
of Monte Carlo steps) at which $f_M(\tau_0,T_p(L))=0.4$ and
$f_E(\tau_0,T_p(L))=0.3$. Here $T_p(L)\rightarrow T_p(4,1)$ for
$L\rightarrow \infty$.
The positive slopes $z_M$ and $z_E$ show that these times diverge in the
thermodynamic limit $L\rightarrow \infty$.
}
\label{corr_M}
\end{figure}

\subsection{The Thermodynamic Transitions}
\noindent

Both the PFF and the PSG model have two thermodynamic transitions,
one associated to the variables $\sigma_i$ and one to the spins $S_i$.
At zero temperature (in 2D) there is a transition for the Ising spins
$S_i$.
The transition is in the IFF universality class for the PFF model, and in the Ising
SG class for the PSG model.
At finite temperature $T_p$ it has been shown, by means of
Monte Carlo (MC) simulations\cite{PSG,PFF} and
analytic approaches\cite{CdLMP}, that 
the Potts variables undergo a transition in
the universality class of a $s$-state Potts ferromagnetic model
(Fig.\ref{lattice}.b). 

It is possible to extend numerically\cite{PFF}
the analytic expression of the ferromagnetic Potts transition
temperature\cite{Wu} to the PSG and the PFF case, by introducing
the fraction $X$ of elementary frustrated cell in the lattice,
with the ferromagnetic Potts model corresponding to $X=0$,
the PSG model to $X=0.5$ and the PFF model  to $X=1$.
The generalized relation is
\begin{equation}
\frac{k_B T_p(q,X)}{a(X)qJ}=\frac{1}{\ln\left[1+\sqrt{a(X)q}\right]} 
\label{TpX=0.5}
\end{equation}
where $q=2s$ and the parameter $a(X)$ is reported in Fig.\ref{lattice}.b.
Note that in 3D the transition of the Ising spins is expected at 
finite $T<T_p$.

\subsection{The Glassy Behavior and the connection with the thermodynamics }
\noindent

It has been shown\cite{breve,franzese0_FFdCC,PFF}  that the PSG and the
PFF models have a glassy behavior.
For example in the PFF with $q\equiv 2s=2,4$ in 2D and 3D
the autocorrelation function
\begin{equation}
f_A(t,T)=\frac{\langle A(t,T) A(0,T)\rangle - \langle A(T) \rangle^2}
{\langle A(0,T)^2 \rangle - \langle A(T) \rangle^2},
\end{equation}
for a global quantity $A$ (like the total energy $E$ or the Potts order
parameters $M$) shows the `two-step' behavior of glasses, with a 
non-exponential second step (upper panel in Fig.\ref{corr_M}.a). 

These global correlation functions 
can be considered as  measures of the structural relaxation
process. 
Their relaxation times ($\tau_0^E$ and $\tau_0^M$ in Fig.\ref{corr_M}.b)
diverge,  in the thermodynamic limit, at the ordering Potts
transition temperature $T_p(q,1)$ in Eq.(\ref{TpX=0.5})\cite{PFF}.
For $T>T_p$ this diverging relaxation time resembles the
$\alpha$-relaxation process occurring for $T>T_g$ in fragile liquids.
\footnote{
For $T<T_p$, these correlation times decrease with $T$ 
(e.g. for $M$ in lower panel in Fig.\ref{corr_M}.a), 
because they are proportional to 
$\xi^z$, where the correlation length $\xi$
is finite for $T\neq T_p$ ($z$ is the quantity-dependent 
dynamical exponent, approximated by the exponents in
Fig.\ref{corr_M}.b). 
In real glasses, instead, the
relaxation time is always increasing for decreasing $T$.
A relaxation time with a more appropriate behavior
 could be the one
associated to the overlap between Potts configurations visited at
different times. We will propose an `extension' of the model, in the
last section, that should give a monotonic slow relaxation time.}

In the PSG case the onset $T^*$ of non-exponentiality coincides with the
Griffiths temperature 
corresponding\cite{Griffiths,breve}  to the transition
temperature $T_c(q)\equiv T_p(q,0)$ of the ferromagnetic $q$-states Potts
model.\footnote{
In disordered systems 
it is possible to show that for finite external field a free
energy (Griffiths) singularity arises\cite{Griffiths}. 
In the limit of external field going to zero, the temperature at which
this singularity occurs goes to the transition temperature $T_c$ of the
corresponding system with no disorder, and the singularity vanishes.
The case with no disorder, corresponding to the $q$-states PSG model, is
the $X=0$ case (i.e. the ferromagnetic $q$-states Potts model) and is
$T_c\equiv T_p(q,0)$.}
This finding generalizes the results for the Ising SG\cite{McMillan,Ogielski}.  
In the Ising SG, indeed, the non-exponential behavior for $T<T_c$ was
related\cite{Randeria} to the 
presence of unfrustrated regions, due to the randomness, with a
size probability distribution that decreases exponentially.
At $T_c$  the ferromagnetic-like correlation length is
equal to the characteristic size of each region, giving rise to a 
non-Gaussian distribution of relaxation times, that is responsible for
the non-exponentiality of the global relaxation time.

Due to the lack of randomness, in the PFF model this explanation is not
valid (the Griffiths temperature is not defined in this case). Indeed,
for the PFF model, the onset of non-exponential relaxation corresponds to
the Potts transition temperature, as has been shown
numerically\cite{PFF,franzese0_FFdCC} 
in 2D for $q=2,4$
(upper panel in Fig.\ref{corr_M}.a) 
and\cite{franzese0_FFdCC}   in 3D for $q=2$.
In these cases the previous argument can be
extended considering that the ferromagnetic-like correlation length is
now associated to the Potts variables, ordering at $T_p$.

Summarizing, the results suggest that if the system is disordered
($X\neq 1$), then $T^*$ 
corresponds to the Potts transition temperature for $X=0$
(i.e. the Griffiths temperature $T_c$); if the
system is fully frustrated ($X=1$), $T^*$ corresponds to 
the Potts transition temperature for $X=1$ (there is no Griffiths
temperature in this case).
Note that the transition is vanishing for $X\neq 1$ (because the 
Griffiths temperature marks a transition that disappears for vanishing
external field) and for the case $X=1$ with $q=2$ 
(in this case the transition is defined only as a
percolation transition, because there are no orientational states).
The transition is actually present for the cases with
$X=1$ and $q>2$.

Hence  the onset of the non-exponentiality is marked by a temperature 
related, at least in this
model, to a thermodynamic transition.
The transition, in this case, is due
to the Potts (orientational)
variables, but, in a more general case, could be associated to any
global (structural) ordering. 
This kind of ordering could be, in
principle, due to some internal degrees of freedom non easily
detected in real experiments.

\section{THE $\beta$-RELAXATION}
\noindent

\begin{figure}
a) \mbox{ \epsfxsize=6.5cm \epsffile{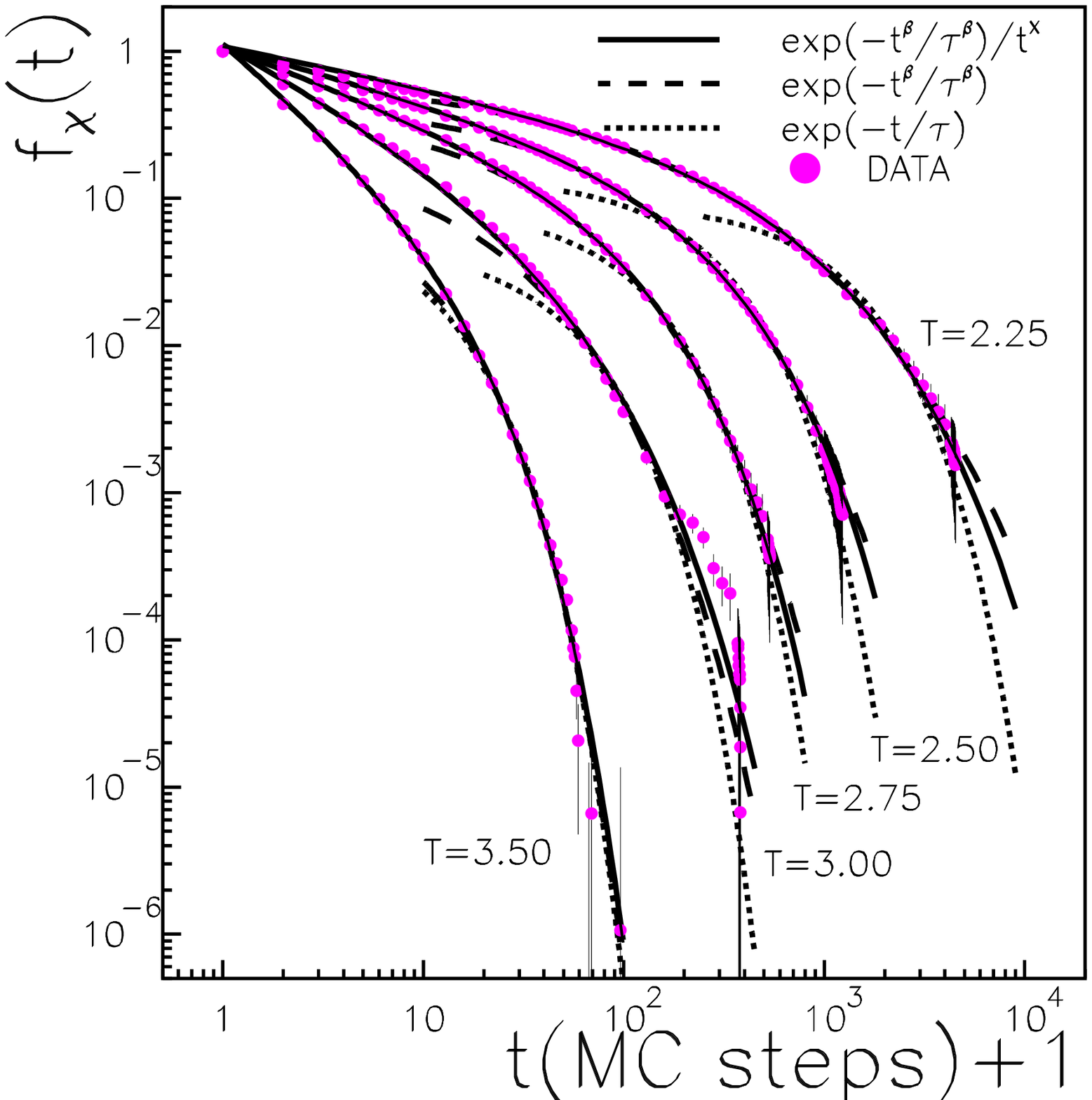} }
b) \mbox{ \epsfxsize=6.5cm \epsffile{ 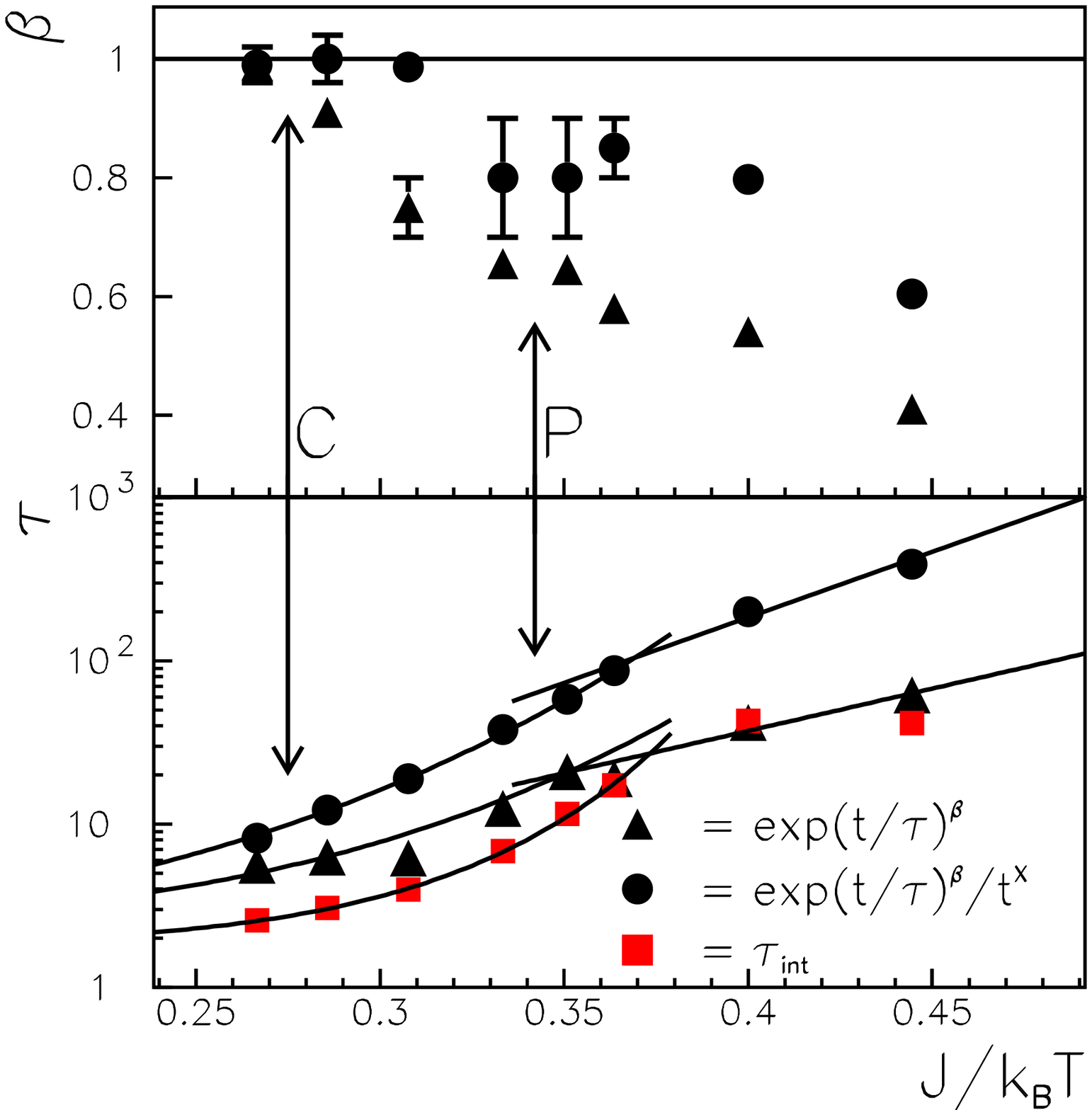  } }
\vspace{0.5cm}
\fcaption{The $\beta$-relaxation in the PSG model in 2D for $q\equiv 2s=4$.
a) The normalized correlation function $f_\chi(t)$ of the nonlinear
susceptibility: temperatures are in units of $J/k_B$; times in units of
MC steps. The data are fitted
with the forms in
Eq.s(\protect\ref{exp},\protect\ref{KWW_stret},\protect\ref{Ogielski_stret}).  
The Ogielski form in Eq.(\protect\ref{Ogielski_stret}) is the only able
to fit the data on 4 decades. Where
not shown the error bars are smaller then the symbols size.
b) The fitting parameters $\beta$ and $\tau$ for the stretched
exponential forms in 
Eq.s(\protect\ref{KWW_stret},\protect\ref{Ogielski_stret}).  
For the values of $x$ see Ref.\protect\cite{breve}. The onset $T^*$ of the
stretched exponential corresponds to the Griffiths temperature $T_c$
(marked with $C$) above the Potts transition temperature $T_p$ (marked
with $P$). 
The lower panel includes also data for the integral
correlation time $\tau_{int}=\lim_{t_{max}\rightarrow
\infty}[\frac{1}{2}+\sum_{t=0}^{t_{max}}f_\chi(t)]$.
The lines are only guides for the eyes.
}
\label{psg}
\end{figure}

Both PSG and PFF models in 2D show\cite{breve} 
 a non-exponential behavior also
for the normalized correlation functions $f_\chi(t)$ 
of the time-dependent nonlinear susceptibility 
\begin{equation}
\chi_{SG}(t)=\frac{1}{N}\overline{\left\langle \left[ \sum_{i=1}^N S_i(t+t_0)
S_i(t_0) \right]^2\right\rangle}
\label{chisg}
\end{equation}
(with $N$ total number of spins, 
$t_0$ equilibration time, $\chi_{SG}(0)=N$)
where the angular brackets stand for the 
thermal average and the bar stands for the average over the disorder in the
PSG case, and is absent in the PFF case. 
Note that $f_{\chi}(t)$ depends explicitly only on the Ising spins.

In the PSG case, the onset $T^*$ of non-exponentiality for $f_\chi$
corresponds to the Griffiths temperature $T_c$ (Fig.\ref{psg}).
In the PFF case it has been shown\cite{breve} that $T^*$ corresponds
to the Potts transition temperature. These results seem to have the same
interpretation as those for the correlation functions of the quantities
depending explicitly on the Potts variables (previous section).

The difference in this case is that 
the Ising spins have no thermodynamic
transition at $T_p$. 
Therefore their correlation time $\tau$, associated to
$f_\chi(t)$, does not
diverge at $T_p$  (lower panel in Fig\ref{psg}.b) 
and, hence, describes a fast process, respect to the slow dynamics 
of the Potts variables.
This fast $\tau$ can be considered as a measure of the
time needed by the Ising spins to minimize the energy locally, 
on the diluted lattice given by the clusters of ordered Potts variables.
This fast (local) relaxation corresponds to the
$\beta$-process.

\begin{figure}
a) \mbox{ \epsfxsize=6.5cm \epsffile{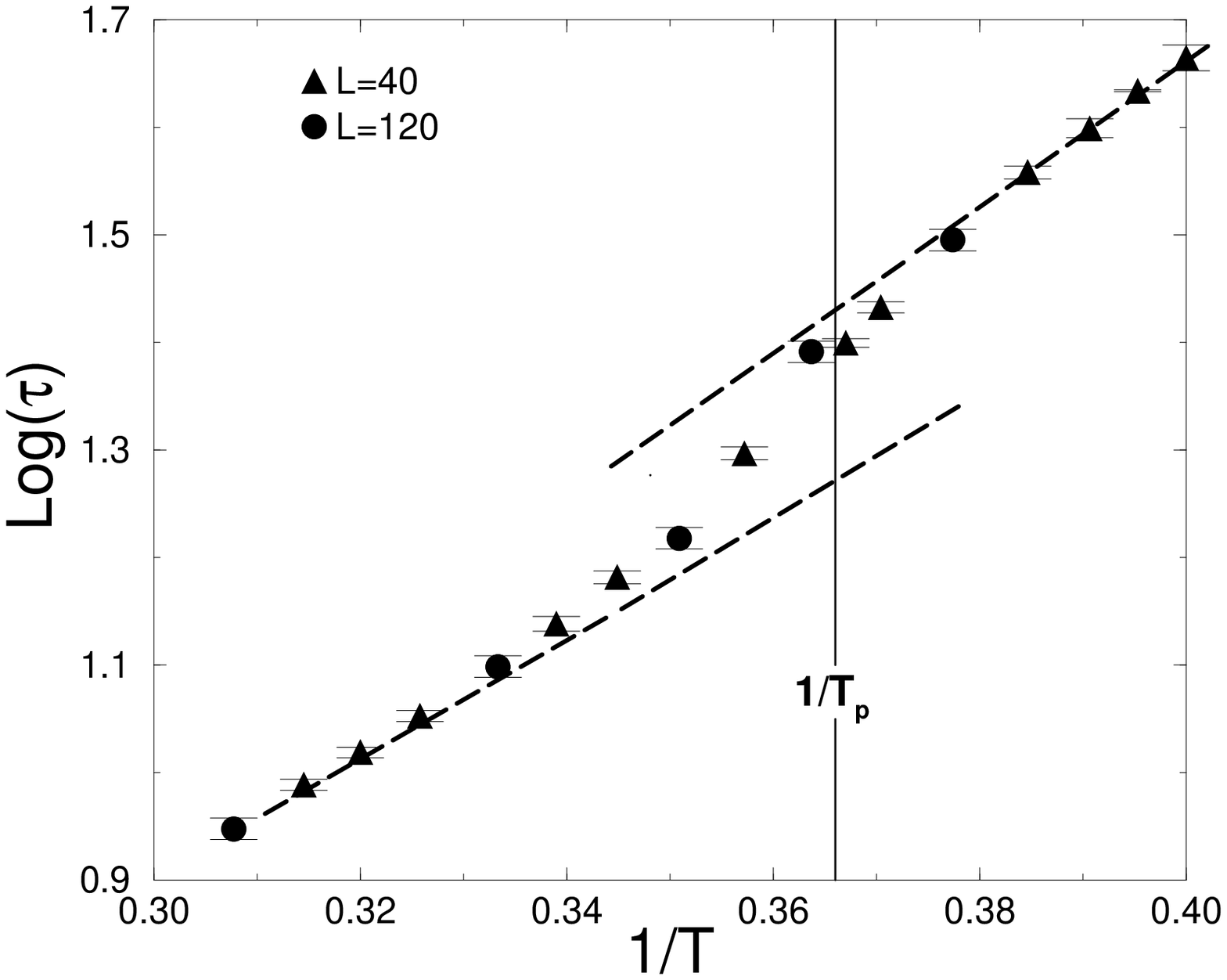} }
b) \mbox{ \epsfxsize=6.5cm \epsffile{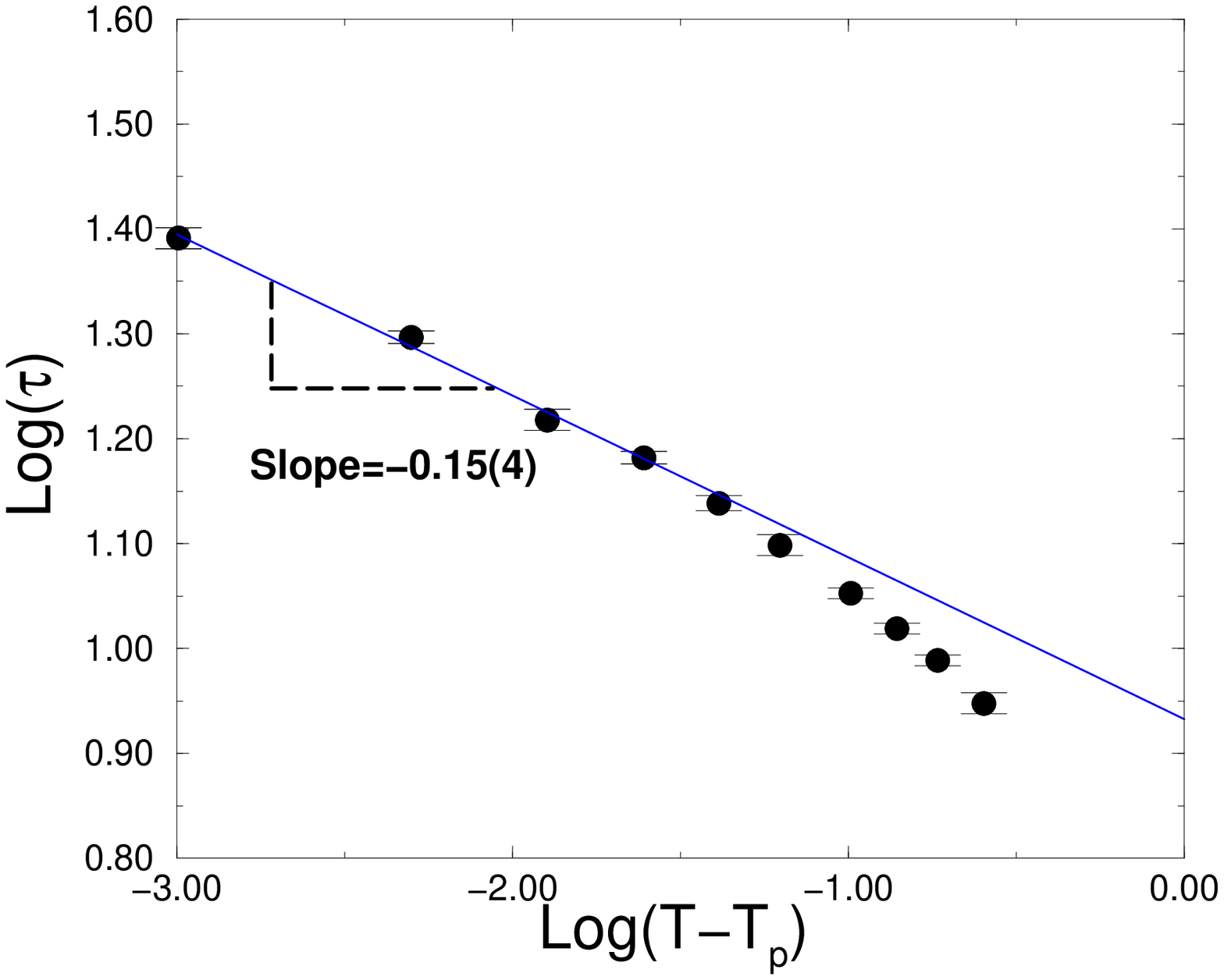} }
\vspace{0.5cm}
\fcaption{The $\beta$-relaxation in the PFF model in 2D with $q=4$.
a) The logarithm of the correlation time $\tau$, associated to $f_\chi$, 
for two finite systems with linear sizes $L=40$, 120, plotted against
the inverse temperature ($T$ is in units of $J/k_B$). The finite size
effect is evident only near the Potts transition temperature $T_p$
(marked by a vertical line),
where the Potts correlation length reach the system size.
An Arrhenius behavior in this plot is represented by a straight line. 
b)
The data for $T>T_p$ plotted against the logarithm of $T-T_p$ to
verify the power law predicted by MCT. The power law is asymptotically
satisfied 
and the slope is an approximate estimate of the power exponent.
In both panels the lines are only guides for the eyes. 
}
\label{pff}
\end{figure}

The intriguing result in this case is that this correlation time $\tau$
has a non-Arrhenius behavior for $T>T_p$ and an Arrhenius behavior for
$T<T_p$ (Fig.\ref{psg}.b and Fig.\ref{pff}).
This result resembles the not well understood 
$\alpha-\beta$ bifurcation, presented in the introduction, that is seen at a
characteristic temperature $T_{\alpha-\beta}$ in some fragile 
liquids\cite{johari} and intermediate liquids\cite{stickel_wagner}.

Moreover the crossover, from non-Arrhenius to Arrhenius,
occurs at $T_p$ 
in both models (with or without disorder).
Therefore $T_{\alpha-\beta}$ is separated by the onset of
non-exponential relaxation function $T^*$ -- marked by $T_c$ if there
is disorder, and by $T_p$ if there is no disorder (previous section).

In the experiments\cite{10q} the temperature $T_{\alpha-\beta}$ seems 
to coincide with the $T_{MCT}$ of the Mode Coupling Theory, at which the
$\alpha$-relaxation time diverges as a power law, Eq.(\ref{MCT}).
Here
the bifurcation 
coincides numerically with
$T_p$, at which the global relaxation
time diverges. 
Indeed, the correlation time $\tau$ can be fitted with the VTF law,
Eq.(\ref{VTF}), with $T_0=T_p$, or can be (asymptotically) well
described by a power law diverging at $T_p$ (e.g. Fig.\ref{pff}.b).

These results suggest that the bifurcation here is
related to the thermodynamic transition of the Potts variables.
The fast $\beta$-process is driven by the slow dynamics of the global
process for $T$ approaching $T_p$ from above. The 
non-Arrhenius behavior of the global process induces a non-Arrhenius
behavior in the coupled fast variable and 
the dominant dynamic process is the one related to the
slow variable. 
As has been shown for supercooled liquids\cite{pes},
is reasonable that in this regime the
dominant dynamic process
is the relaxation to the lower 
accessible energy, at the given $T$, following the 
instability directions, in the energy landscape, 
 of the visited states, i.e. the  activated
processes play no role.

For the supercooled liquid case\cite{pes}
it has been shown that $T_{MCT}$ corresponds to the temperature below
which the average number of instability directions for the visited
configuration goes to zero. 
From preliminary results, 
the same situation appears to be valid
here at $T_p$.

Finally,
below $T_p$ the system is exploring one of the basins of attraction of
the Potts variables $\sigma_i$
 and the dynamics of the fast Ising spins is no longer
coupled to that of the $\sigma_i$. The dynamics of the fast variables
become activated (Arrhenius) consistently with
the lack of instability directions. Indeed, in this case the energy
cannot be lowered by following an instability direction, but only by
doing an activated process.
This is analogous of what has been found for the global relaxation 
in supercooled liquids\cite{pes}, but in our case the
focus is on the local $\beta$-processes.

Therefore the results here suggest that the
$\alpha-\beta$ bifurcation can be related to the thermodynamics of the
systems, and that the Potts transition temperature $T_p$ of these
systems plays the role of the $T_{MCT}$ of supercooled liquids.

\section{CONCLUSION AND PERSPECTIVES}
\noindent

The glassy spin models considered here, the PSG and the PFF model,
are characterized by the presence of two coupled variables, whose
dynamics decouple when the relaxation time of the slow variable
diverges.
This happens at the temperature $T_p$ where the slow variable has 
a thermodynamic phase transition.

The picture that can be derived is the following.
Due to the symmetry breaking of the slow variable for $T\leq T_p$, the
system is attracted in one of the basin of the energy landscape of the
slow variable. 
Inside this basin, the dynamics of the system is mainly due to the fast
variable and the dominant dynamical mechanism is the
activated process, as consequence of the lack of instability directions
for the visited states. 

Macroscopically, this is shown by the $\alpha$-$\beta$ bifurcation that
can be seen in some fragile and intermediate glassforming 
liquids\cite{johari,10q}, i.e. by the crossover of the $\beta$-relaxation time
from a non-Arrhenius behavior to an Arrhenius law at a temperature
$T_{\alpha-\beta}$. 

The experiments suggest that $T_{\alpha-\beta}\simeq T_{MCT}$,
consistently with the analysis reported here. Indeed, in our models
the bifurcation occurs at the temperature which plays the role of $T_{MCT}$,
i.e.  $T_p$, where the 
slow process has a diverging relaxation time.

Therefore the $\alpha$-$\beta$ bifurcation can been related to the
thermodynamics, as well as the onset $T^*$ of non-exponential relaxation.
What we can learn from the `toy model' studied here is that, in cases in which there is
disorder, $T^*$ and $T_{\alpha-\beta}$ do not coincide. The first,
indeed, corresponds to the Griffiths temperature $T_c$ 
-- associated to a real transition only for a non-zero
external field coupled to the slow variable -- and the
second to 
the temperature $T_p$, where the slow relaxation time diverges.
The two temperatures coincides only in the particular case of a
frustrated system with no disorder.

An open questions is the effect of
the $T$-dependence of the global correlation length $\xi$.
An interesting case is the
one in which $\xi$ 
does not decrease below the thermodynamic
transition temperature. A model with such characteristic is the $XY$
model, undergoing the Kosterlitz-Thouless-Berezinskii\cite{BKT}
transition.
The resulting Hamiltonian will be
\begin{equation}
H\{S_i,\phi_i,\epsilon_{i,j}\}
=-J\sum_{\langle i,j \rangle} \cos(\phi_i-\phi_j-A_{i,j})
(\epsilon_{i,j}S_i S_j+1)
\end{equation}
where $A_{i,j}$ are constants depending on the gauge (or the external
field) and 
$\phi_i\in [0,2\pi)$ are a more realistic representation of the 
continuous orientational variables.

\pagebreak

\section{ACKNOWLEDGMENTS}
\noindent
This work is consequence of  many years of collaboration with Antonio
Coniglio, during my studies in Naples where I benefited from his advise and
his friendship. 
It has been an honor for me to be his student.
I wish to thank C. Austen Angell for bringing to my attention the
$\alpha$-$\beta$ bifurcation  in a very instructive private
discussion. 
I am in debt with S. Franz, G. Parisi, S. Mossa, E. La Nave, G. Ruocco,
A. Scala, F. Sciortino, H.E. Stanley, F. Starr for enlightening
discussions and  precious advises.


\section{REFERENCES}
\noindent
\begin{thebibliography}{000}
\bibitem{10q}
C. A. Angell, {\tenit J. Phys. C} {\tenbf 12} (2000) 6463.
\bibitem{Young}
See for example, {\tenit Spin Glasses and Random Fields}, ed. A. P. Young
(World Scientific, Singapore, 1998).
\bibitem{CdLMP}
A. Coniglio, F. di Liberto, G. Monroy, and F. Peruggi, {\tenit
Phys. Rev. B} {\tenbf 44}, 12605 (1991).
U. Pezzella and A. Coniglio, {\tenit Physica A} {\tenbf 237}, 353 (1997).
F. di Liberto and F. Peruggi, {\tenit Physica A} {\tenbf 248}, 273 (1998).
\bibitem{PSG}
G. Franzese and A. Coniglio, {\tenit Phys. Rev. E} {\tenbf 58}, 2753 (1998).
\bibitem{franzese0_FFdCC}
G. Franzese, A. Fierro, A. De Candia, and A. Coniglio, {\tenit Physica
A} {\tenbf 257}, 376 (1998).
A. Fierro, G. Franzese, A. de Candia, and A. Coniglio, {\tenit Phys. Rev. E}
{\tenbf 59}, 60 (1999). 
\bibitem{breve}
G. Franzese and A. Coniglio, {\tenit Phys. Rev. E.} {\tenbf 59}, 6409 (1999). 
\bibitem{phil}
G. Franzese and A. Coniglio, {\tenit Philos. Mag. B} {\tenbf 79}, 1807 (1999).
\bibitem{PFF}
G. Franzese, {\tenit Phys. Rev. E.} {\tenbf  61}, 6383 (2000). 
\bibitem{johari}
G. P. Johari and M. Goldstein, {\tenit J. Chem. Phys.} {\tenbf 53}, 2372
(1970).
G. P. Johari and M. Goldstein, {\tenit J. Chem. Phys.} {\tenbf 55}, 42459
(1971).
\bibitem{stickel_wagner}
F. Stickel, E. W. Fisher, and R. Richert, {\tenit J. Chem. Phys.}
{\tenbf 102}, 6251 (1995).
C. Hansen, F. Stickel, T. Berger, R. Richert, and E. W. Fisher, {\tenit
J. Chem. Phys.} {\tenbf 107}, 1086 (1997).
H. Wagner and R. Richert, {\tenit J. Phys. C} {\tenbf 103} 4071 (1999).
\bibitem{pes}
E. La Nave, A. Scala A, F. W. Starr, F. Sciortino, and H. E. Stanley {\tenit
Phys. Rev. Lett.} {\tenbf 84}, 4605 (2000).
A. Scala, F. W. Starr, E. La Nave, F. Sciortino, and H. E. Stanley, 
{\tenit Nature} {\tenbf 406}, 166 (2000).
L. Angelani, R. Di Leonardo, G. Ruocco, A. Scala, and F. Sciortino, {\tenit
Phys. Rev. Lett.} {\tenbf 85}, 5356 (2000).
S. Satry, {\tenit Nature} {\tenbf 409}, 164 (2001).
L. Angelani, R. Di Leonardo, G. Parisi, G. Ruocco, preprint cond-mat/0011519.
\bibitem{Angell}
C. A. Angell, in {\tenit Relaxation in Complex Systems}, ed.  K. L. Ngai and 
G. B. Wright (US Department of Commerce, Springfield, 1985).
K. Ito, C. T. Moynihan, and C. A. Angell {\tenit Nature} {\tenbf 398}, 492 (1999).
\bibitem{Gotze}
W. G\"otze, in {\tenit Liquids, freezing and glass transition}, 
ed. J. P. Hansen, D. Levesque, J. Zinn-Justin (North Holland,
Amsterdam, 1991).
W. G\"otze and L. Sj\"ogren {\tenit Rep. Prog. Phys.} {\tenbf 55}, 241 (1992).
\bibitem{Ogielski}
A. T. Ogielski, {\tenit Phys. Rev. B} {\tenbf 32}, 7384 (1985).   
\bibitem{mossa}
G. Monaco, S. Caponi, R. Di Leonardo, D. Fioretto and G. Ruocco,
{\tenit Phys. Rev. E} {\tenbf 62}, 612 (2000).
S. Mossa et al. unpublished.
\bibitem{Wu} 
F. Wu, {\tenit Rev. Mod. Phys.} {\tenbf 54}, 235 (1982). 
\bibitem{Griffiths}
R. B. Griffiths, {\tenit Phys. Rev. Lett.}, {\tenbf 23}, 17 (1969).
\bibitem{McMillan}
W. L. McMillan, {\tenit Phys. Rev. B} {\tenbf 28}, 5216 (1983).
\bibitem{Randeria}
M. Randeria, J. P. Sethna, and R. G. Palmer, {\tenit Phys. Rev. Lett.} {\tenbf 57}, 245 
(1986). 
A. Bray, {\tenit Phys Rev. Lett.} {\tenbf 59}, 586 (1987). 
F. Cesi, C. Maes, and F. Martinelli, {\tenit Commun. Math. Phys.} {\tenbf 188},
135 (1997).
\bibitem{BKT}
    	V. L. Berezinskii, {\tenit Zh. Eksp. Teor. Fiz.} {\tenbf 59}, 207
    	(1970); {\tenit Sov. Phys. JETP} {\tenbf 32}, 493 (1971)].
    	J. M. Kosterlitz and D. J. Thouless, {\tenit J. Phys. C} {\tenbf 6},
    	1181 (1973);

\end{thebibliography}
\end{document}